\documentclass[prl,twocolumn,showpacs,aps]{revtex4}
\usepackage{graphics}
\usepackage{dcolumn}
\usepackage[sort&compress]{natbib}
\newcommand{\ab}{{\bf a}}

\newcommand{\eb}{{\bf e}}
\newcommand{\pbb}{{\bf p}}
\newcommand{\rb}{{\bf r}}

\newcommand{\Eb}{{\bf E}}
\newcommand{\Pb}{{\bf P}}
\newcommand{\Rb}{{\bf R}}

\newcommand{\Ic}{{\cal I}}

\newcommand{\Pc}{{\cal P}}
\newcommand{\tp}{\tilde{p}}
\newcommand{\tq}{\tilde{q}}
\newcommand{\Arr}{\mathop{\longrightarrow}}
\newcommand{\mod}{\mathop{\mathrm{mod}}\nolimits}
\begin{document}
\title{Pyroelectric vector and related polarization effects}
\author{Eugene V. Kholopov}
\email{kholopov@casper.che.nsk.su}
\affiliation{Institute of Inorganic Chemistry of the Siberian Branch of
the Russian Academy of Sciences, 630090 Novosibirsk, Russia}
\begin{abstract}
The principal fact that the pyroelectric vector $\Pb^0$ typical of polar
crystals is invisible in equilibrium has been verified. Nevertheless, the
general formula for $\Pb^0$ as a definite bulk quantity associated only
with the inversion properties of a given charge distribution in the bulk
has been obtained. Its application to ferroelectric perovskites BaTiO$_3$,
PbTiO$_3$, and KNbO$_3$ is discussed.
\end{abstract}
\maketitle

For many years the pyroelectric vector stays to be one of the most
mysterious objects in solid state physics, though the phenomena of pyro-
and ferroelectricity are classical \cite{Cady46,Line77}. The attempts to
grasp it result in the concept that this value can be described within an
adiabatic evolution between characteristic states \cite{Aizu62}. The
elegant realization of this approach for electron subsystems \cite{King93}
is based on the formalism of the Berry phase \cite{Berr84} in which case
the integration over a closed contour in a parametric space arises and is
reduced to a quantized flow through that contour. Many important results
are obtained within this approach \cite{Rest94,Zho94a,Zho94b,Mart97}.
However, an evolution route over equilibrium states \cite{Aizu62,Berr84}
is a subtle problem characteristic of thermodynamic charge systems
\cite{King93}, as discussed below. One more problem is that the
pyroelectricity is a structural property of polar crystals
\cite{Cady46,Land84} and so must be typical of every crystal structure
without inversion, notwithstanding is the description classical or quantum.
It implies that apart from the quantum Berry-phase approach
\cite{Rest94,Orti94}, other more classical treatments should exist.

Here we propose a novel approach to this subject and represent the
pyroelectric vector $\Pb^0$ in the form of a general invariant specifying
polar crystals. Expressed in terms of characteristic length parameters and
effective charges in the bulk, $\Pb^0$ is defined as a function of state,
without recourse to any evolution. As a result, it can be estimated with
making use of classical charge distributions, though a quantum-mechanical
calculation of effective charges is eventually assumed.

Interested in periodic charge distributions in the bulk, we first point
out that any local polarization $\Pb$ as a bulk property associated with
the crystal symmetry \cite{Cady46,Land84,Mart72} and so independent of a
small external field $\Eb$ should give the field contribution $-\Pb\Eb$ to
the specific energy, but this contribution is absent in equilibrium. To
visualize this fact, we note that the projection $P_z$ along any principal
axis $z$ is determined by the effective one-dimensional charge
distribution $\tilde{\rho}(z)=\tilde{\rho}(z+c)$, with the period $c$
appropriate to the case, which arises after integrating the initial charge
distribution over normal cross-sections of a unit cell. The value of
$P_z$ can then be written down as
\begin{equation}\label{Eq1}
P_z=\frac{1}{c}\int_0^cdg\Bigl[\int_g^c z\tilde{\rho}(z)dz
+\int_0^g(z+c)\tilde{\rho}(z)dz\Bigr]=0 .
\end{equation}
Here the fact that the uniform field component $E_z$ does not distinguish
any fixed choice of the unit cell boundaries in the bulk is taken into
account. So, the expression in square brackets describes the polarization
of the unit-cell portion of $\tilde{\rho}(z)$ shifted by $g$ from any
initial one, while the outer integral with the factor $1/c$ accounts for
the averaging over all shifts as equiprobable. After interchanging the
order of integration in (\ref{Eq1}), the final result follows from charge
neutrality of a unit cell.

This issue follows the statement by Larmor \cite{Cady46,Larm21} that
any bulk dipolar properties are not typical of crystals in equilibrium.
Such states are associated with the thermodynamic limit
\cite{Onsa39,Grif68}, with sample surfaces adjusted so as to exclude
depolarization effects \cite{Aizu62,Smit81}. Notwithstanding, if a finite
sample is metastable because of unstable charged surface \cite{Task79},
the surface charges always form a macroscopic dipole which interacts with
an external field and, being proportional to a sample volume, can be
formally represented as an additive composition of dipoles in the bulk.
Such dipoles have, without doubt, the bulk nature at the first moment of
forming a steady-state surface \cite{King93,Land84}. But due to the
disturbance of a bulk symmetry near the surface, initial polarization
effects do not recognize polar crystals \cite{Larm21,Land81}. To single
out true bulk effects, the scheme of a sample placed between shorted
condenser plates normal to the polar axis is usually applied
\cite{Cady46,Line77}. But the effect of $\Pb^0$ is then also removed in
accord with (\ref{Eq1}), so that only increments in $\Pb^0$ caused by
changes in temperature or stresses are actually recorded via transient
signals \cite{Cady46}.

Since $\Pb^0$ is evanescent for observation on polar crystals in
equilibrium \cite{Cady46,Land84,Ginz46} and the polarity reversal is
impossible as an equilibrium physical process in pure pyroelectrics, the
definition of $\Pb^0$ as a bulk measure of polarity associated with the
inversion symmetry of a given state alone should be addressed. To gain
insight into this subject, we first consider a simple Bravais lattice
composed of point charges $q$. We assume that this lattice is immersed in
a uniform neutralizing charge background and are interested in a dipolar
moment $\pbb$ of the volume of a unit-cell parallelepiped specified by
elementary translations $\ab_1$, $\ab_2$, and $\ab_3$ and situated anyhow
in the bulk. If the inner point charge is displaced by $\rb$ from the
parallelepiped center initially, then it is easy to show that upon any
displacement $\Rb$ of that parallelepiped through the bulk, with casting
$\rb-\Rb$ in terms of $\{\ab_k\}$, $\pbb(\Rb)$ takes the form
\begin{equation}\label{Eq2}
\pbb(\Rb)=q\Rb_0 ,\quad q(\rb-\Rb)\equiv q\Rb_0(\mod\,\{q\ab_k\}) ,
\end{equation}
provided that $\Rb_0$ belongs to the parallelepiped volume. This result is
evident, for when a recurrent point charge leaves the unit-cell volume,
the inner portion of background begins to neutralize the next point charge
entering. So, $\pbb(\Rb)$ is a periodic saw-tooth function of $\Rb$ with
vertical jumps and with zero mean value in agreement with (\ref{Eq1}).
Sites of this Bravais lattice and midpoints in between are its inversion
points. Thus, a unit cell centered on any of those inversion points is
described by $\pbb=0$. In the cases of midpoints it implies that any of
$q$ on the unit-cell boundaries is to be evenly shared between adjacent
unit cells \cite{Aizu62,Evje32}, in accord with (\ref{Eq2}).

Every regular structure of point charges of different species may be
cast in terms of simple Bravais sublattices introduced. To describe
their polarization, the one-dimensional treatment is expedient along
directions normal to each couple of parallel faces of the foregoing
elementary parallelepiped so as to exclude multiple projections of every
particular Bravais sublattice on the axis under examination. These
directions are defined by the unit vectors $\eb_k$ and by the periods
$d_k=(\ab_k\eb_k)$, where
\begin{equation}\label{Eq3}
\eb_1=\frac{\ab_2\times\ab_3}{[|\ab_2|^2|\ab_3|^2
-(\ab_2\ab_3)^2]^{1/2}}
\end{equation}
and the transformation $\eb_1\to\eb_2\to\eb_3$ is cyclic.

\begin{figure}[t]
\resizebox{0.95\hsize}{!}{\includegraphics{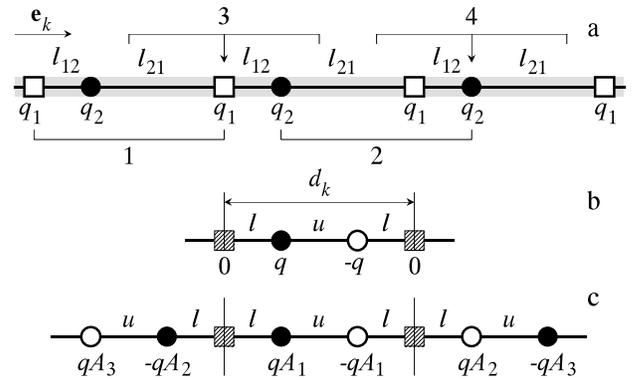}}
\caption{(a) Linear structure of two point-charge sublattices in a
compensating uniform background shaded, with indicating the symmetric unit
cell choices. (b) Unit cell '1' after removing the symmetric part of
charges by means of inversion relative to its center, with $q=q_2/2$. (c)
The further modification of unit cell '1' allowing for the inversion
antisymmetry relative to the unit-cell ends.}\label{Fig1}
\end{figure}

We begin with a combination of two Bravais sublattices, which is then
specified by their projections per normal section of an elementary
parallelepiped on a given $k$-direction, as shown in Fig.~\ref{Fig1}(a),
where the label $k$ on all the parameters is omitted for brevity. The
foregoing four symmetrical choices of the unit cell lead to four different
values of the dipolar moment:
\begin{eqnarray}
&{\displaystyle p_1=\frac{(l_{12}-l_{21})\,q_2}{2}\, ,\quad
p_2=-\frac{(l_{12}-l_{21})\,q_1}{2}\, ,}& \label{Eq4}\\
&\cases{{\displaystyle\hspace{0.2em} p_3=\phantom{-}l_{12}\,q_2 ,\quad
p_4=-l_{12}\,q_1}& at $\hspace{0.2em} l_{12}<l_{21}$,\cr
{\displaystyle\hspace{0.2em} p_3=-l_{21}\,q_2 ,\quad
p_4=\phantom{-}l_{21}\,q_1}& at $\hspace{0.2em} l_{12}>l_{21}$.}&
\label{Eq5}
\end{eqnarray}
Also keeping in mind the following combination
\begin{equation}\label{Eq6}
\frac{p_1+p_2}{2}=-\frac{(l_{12}-l_{21})(q_1-q_2)}{4}
\end{equation}
antisymmetric to inversion, we conclude that the general antisymmetric
invariant containing all the motifs typical of (\ref{Eq4})--(\ref{Eq6})
takes the form
\begin{equation}\label{Eq7}
\Ic(q_1,q_2;l_{12},l_{21})=l_{12}l_{21}(l_{12}-l_{21})q_1q_2(q_1-q_2) .
\end{equation}
Each of six zeros of (\ref{Eq7}) excludes any polarization and this set of
zeros is complete. The symmetry of (\ref{Eq7}) also implies that in the
case at hand the inversion can be defined in the four equivalent manners:
(i) $l_{12}\to-l_{12}$ and $l_{21}\to-l_{21}$; (ii) $q_1\to-q_1$ and
$q_2\to-q_2$; (iii) $l_{12}\leftrightarrow l_{21}$; (iv)
$q_1\leftrightarrow q_2$. Note that the transformation
$(q_1;l_{12})\leftrightarrow(q_2;l_{21})$, that is the combination of
(iii) and (iv), results in $p_1\leftrightarrow p_2$, but $p_3\to p_4$
takes place after the additional shift $\mp q_1(l_{12}+l_{21})$, where
$l_{12}+l_{21}=d_k$ in agreement with (\ref{Eq2}). Hence, those couples of
cells are not mixed by operations related to inversion and so may be
discussed separately.

To create (\ref{Eq7}) normalized per unit cell, we start from unit cell
'1', where the antisymmetric charge combination arises as a result of
inversion relative to the unit-cell midpoint, as shown in Fig.
\ref{Fig1}(b). The end charges and a background then disappear, so that
the constraint of neutrality becomes explicit, with conserving $p_1$ in
form (\ref{Eq4}). Next we decompose each charge above into fractions
$A_1$, $A_2$, and $A_3$, which are then inverted relative to the ends of
the unit cell in an antisymmetric fashion shown in Fig. \ref{Fig1}(c).
Taking into account the relation $A_1+A_2+A_3=1$ normalizing the result
per actual charges, the dipolar moment $\tp_1$ of modified cell '1' can be
written as
\begin{equation}\label{Eq8}
\tp_1=q_2\Bigl[\frac{l_{12}-l_{21}}{2}+d_kB_1\Bigr]=p_1+q_2d_kB_1 ,
\end{equation}
where $B_1=A_2-A_3$ is a new parameter. Thus the effect of inversion
is realized in form (\ref{Eq8}) as a shift of the $q_2$-sublattice
relative to the $q_1$-one.

The value of $\tp_2$ is defined in the same way, but with $B_2=-B_1$ that
supports the above treatment in terms of shifts and follows from the
transformation $\tp_1\leftrightarrow\tp_2$ associated with
$(q_1;l_{12})\leftrightarrow(q_2;l_{21})$. Next we assume a functional
form $B_j=D_j/D_0$ so as to satisfy (\ref{Eq7}), where $D_j$ are
polynomials bilinear in $q_j$ and $l_{jj'}$, provided that $D_1=-D_2$ is
antisymmetric upon $(q_1;l_{12})\leftrightarrow(q_2;l_{21})$, whereas
$D_0$ is symmetric upon both $q_1\leftrightarrow q_2$ and
$l_{12}\leftrightarrow l_{21}$. Then the quantity of interest is of the
form
\begin{eqnarray}
&{\displaystyle\Pc_2(q_1,q_2;l_{},l_{})=\frac{\tp_1+\tp_2}{2}=-\frac{t\,
\Ic(q_1,q_2;l_{12},l_{21})}{2D}} ,&\label{Eq9}\\
&D=d_k^2\bigl(q_1^2+q_2^2+t_1q_1q_2\bigr)+2tl_{12}l_{21}q_1q_2 ,&
\label{Eq10}
\end{eqnarray}
with free parameters $t$ and $t_1$. Likewise, the same result arises upon
considering cells '3' and '4', though $B_3=-B_4$ differs from $B_1$
therein. It implies that the solution (\ref{Eq9})--(\ref{Eq10}) is a
mutual property of two sublattices, but not of their particular unit-cell
representations.

Here we tacitly imply that both $q_1$ and $q_2$ are of the same sign and
so homogeneous. In this case form (\ref{Eq7}) is also antisymmetric with
respect to the replacement: $q_1\to l_{12}$, $l_{12}\to q_2$, $q_2\to
l_{21}$, and $l_{21}\to q_1$. In the event of (\ref{Eq9}) and (\ref{Eq10})
this fact results in $t_1=2$.

If $(q_1q_2)<0$, a similar consideration, but with $|q_1q_2|$ instead of
$q_1q_2$ in $D_j$, gives rise to (\ref{Eq9}), where
\begin{equation}\label{Eq11}
D=d_k^2\bigl(q_1^2+q_2^2+t_2|q_1q_2|\bigr)+2tl_{12}l_{21}|q_1q_2| .
\end{equation}
In this case there is an additional limiting condition
\begin{equation}\label{Eq12}
\Pc_2\{q,-q;a,b\}\Arr_{b\to\infty}-aq
\end{equation}
corresponding to a Bravais lattice of point dipoles with a period $b$
and leading to $t_2=2-t$. The parameter $t$ is common to (\ref{Eq10}) and
(\ref{Eq11}) from the continuity of (\ref{Eq9}). On comparing the limit of
(\ref{Eq9}) as $q_1\to\infty$ and $l_{12}\to\infty$ to (\ref{Eq5}), we
obtain $t=2$, because $q_2$ and $l_{21}$ then describe a small perturbation.
Note that the other three choices of limiting couples of parameters are also
admissible. As a result, we arrive at the normalized value
\begin{eqnarray}\label{Eq13}
\Pc_2(q_1,q_2;l_{12},l_{21})&&\nonumber\\
&&\hspace{-7em}{}=\cases{{\displaystyle\frac{-\Ic(q_1,q_2;l_{12},l_{21})}{d_k^2
(q_1+q_2)^2+4l_{12}l_{21}q_1q_2 }}& at $(q_1q_2)>0$,\cr
{\displaystyle\frac{-\Ic(q_1,q_2;l_{12},l_{21})}{d_k^2(q_1^2+q_2^2)
-4l_{12}l_{21}q_1q_2}}& at $(q_1q_2)<0$.\qquad}
\end{eqnarray}

The treatment above can be generalized to $n$ sublattices described by an
ordered set $\{q_j\}=\{q_1,\dots,q_n\}$ and by a complementary set
$\{l_{jj+1}\}=\{l_{12},l_{23},\dots,l_{n1}\}$ of nearest-neighbor
distances. Keeping in mind that every couple of charges gives the
independent contribution described by (\ref{Eq13}), we readily reach
\begin{equation}\label{Eq14}
\Pc_n(\{q_j\};\{l_{jj+1}\})=\frac{1}{n}\sum_{j=1}^n\sum_{j'\neq j}
\Pc_2(q_j,q_{j'};l_{jj'},l_{j'j}) ,
\end{equation}
where $l_{jj'}+l_{j'j}=d_k$ for every $j$ and $j'$. Every particular
$\Pc_2(q_j,q_{j'};l_{jj'},l_{j'j})$ is counted twice in (\ref{Eq14}),
so that its own normalization is neutralized there.

It is important that $\Pc_n(\{q_j\};\{l_{jj+1}\})$ is not additive to
the formal splitting of any nonzero $q_j$ into fractions, of which number
would multiply the contribution of $\tp_j$ to $\Pc_n(\{q_j\};\{l_{jj+1}\})$,
with destroying the symmetry effect. Thus, every set of coinciding
projections of point lattices must be reckoned once.

To describe a real structure composed of $m$ sublattices of point nuclei
$q_j^{\mathrm{nuc}}$ with a compensating electron density in the above
manner, we map $q_j^{\mathrm{nuc}}$ onto the sets $\{q_{k,j}\}$ of $n_k$
point charges, with retrieving the label $k$ associated with $\eb_k$,
whereas the three electron charge projections $\rho_k^{\mathrm{el}}(z)$
on these $k$-directions are defined like $\tilde{\rho}(z)$ in (\ref{Eq1}).
Since any of $\rho_k^{\mathrm{el}}(z)$ is consistent with the structural
parameters $\{l_{jj+1}^{(k)}\}$ connected with $\{q_{k,j}\}$ in equilibrium,
its effect on (\ref{Eq14}) is reduced to the replacement of $q_{k,j}$ by
effective charges:
\begin{equation}\label{Eq15}
\tq_{k,j}=q_{k,j}+q_{k,j}^{\mathrm{el}} ,\quad
Q=\sum_{j=1}^{n_k}q_{k,j}^{\mathrm{el}}=-\sum_{j=1}^{m}q_j^{\mathrm{nuc}},
\end{equation}
where $Q$, the sum of $q_{k,j}^{\mathrm{el}}$, is determined by the last
equality allowing for the total electrical neutrality.

To determine $q_{k,j}^{\mathrm{el}}$, a new set of unit cells based on
the set $\{l_{jj+1}^{(k)}\}$ is to be introduced so as to avoid the
degeneracy associated with symmetric charge positions used above. To this
end, we note that any of the intervals $l_{jj+1}^{(k)}$ is topologically
equivalent, notwithstanding is it large or small relative to the others,
and so must contain the boundary points of just one unit-cell species $j$
of length $d_k$. By homogeneity, the boundaries of the $j$th unit cells are
assumed to be shifted by $\xi_kl_{jj+1}^{(k)}$ from the centers of the intervals
$l_{jj+1}^{(k)}$ at a given $j$, with $-1/2<\xi_k<1/2$ common to all $j$.
The values of $q_{k,j}^{\mathrm{el}}$ are then determined by the set of
$n_k$ linear equations which, at different $j$, describe the dipolar moments
$F_j^{(k)}(\xi_k)$ of the electron charge distributions within the $j$th unit
cells in terms of the contributions of point charges $q_{k,j'}^{\mathrm{el}}$
located properly:
\begin{eqnarray}
&{\displaystyle\sum_{j'=1}^{n_k}q_{k,j'}^{\mathrm{el}}r_{jj'}^{(k)}=
F_j^{(k)}(\xi_k)} ,&\label{Eq16}\\
&{\displaystyle F_j^{(k)}(\xi_k)=\int_{-d_k/2}^{d_k/2}z'\rho_j^{(k)}
(\xi_k,z')\,dz'=QR_j^{(k)}(\xi_k)} .\quad&\label{Eq17}
\end{eqnarray}
Here the center of the $j$th unit cell is the origin common to the
position $r_{jj'}^{(k)}$ of the $j'$th charge, to the variable $z'$
specifying $\rho_j^{(k)}(\xi_k,z')$ from $\rho_k^{\mathrm{el}}(z)$, and to
$R_j^{(k)}(\xi_k)$, the center of gravity of the electron charge in that
unit cell. On substituting the right equality of (\ref{Eq17}) into
(\ref{Eq16}), with taking (\ref{Eq15}) into account, the equations
(\ref{Eq16}) become homogeneous, so that their nontrivial solution exists
at
\begin{equation}\label{Eq18}
\det\left|r_{jj'}^{(k)}-R_j^{(k)}(\xi_k)\right|=0 .
\end{equation}
This is quite natural, because $Q$ furnishes one more constraint on
$q_{k,j'}^{\mathrm{el}}$. On casting $r_{jj'}^{(k)}$ in terms of
$l_{jj+1}^{(k)}$, one can prove by induction that relation (\ref{Eq18}) is
converted into
\begin{equation}\label{Eq19}
\sum_{j=1}^{n_k}l_{jj+1}^{(k)}\Bigl[F_j^{(k)}(\xi_k)
+\xi_kQ\,l_{jj+1}^{(k)}\Bigr]=0 .
\end{equation}
With making use of the left equality of (\ref{Eq17}), equation
(\ref{Eq19}) specifies a self-consistent value of $\xi_k$, while set
(\ref{Eq16}) gives rise to $q_{k,j'}^{\mathrm{el}}$, so that $\tq_{k,j}$
in (\ref{Eq15}) has been definite.

On joining all the results obtained together, the pyroelectric density
vector is well-defined by the form
\begin{equation}\label{Eq20}
\Pb^0=\frac{1}{v}\sum_{k=1}^3\eb_k\Pc_{n_k}(\{\tq_{k,j}\};
\{l_{jj+1}^{(k)}\}) ,
\end{equation}
where $v$ is the unit-cell volume. Deduced from pristine charges
by linear operations, (\ref{Eq20}) agrees with (\ref{Eq1}).

Note that in the phase-transition problem relation (\ref{Eq20}) implies
that a sole nonzero component of $\Pb^0$ can arise without jumps,
\begin{table}[t]
\caption{Pyroelectric vector and concomitant parameters of some
ABO$_3$ perovskites which are ferroelectric in a tetragonal phase,
with model uniform spherical electron distributions employed.
Experimental and Berry-phase-approach (BPA) results are also given for
comparison.}\label{Table1}
\begin{ruledtabular}
\begin{tabular}{lddd}
&$BaTiO$_3$\hspace{-2.7em}$&$PbTiO$_3$\hspace{-2.7em}$&$KNbO$_3
$\hspace{-2.5em}$\\
\hline \vspace{-3mm}\\
$a$ (nm)\tablenotemark[1]&0.39947&0.3904&0.3997 \\
$c$ (nm)\tablenotemark[1]&0.40336&0.4150&0.4063 \\
$\eta_{\rm{B}}$\tablenotemark[1] &0.012&0.041&0.019 \\
$\eta_{\rm{I}}$\tablenotemark[1] &-0.023&0.112&-0.025 \\
$\eta_{\rm{II}}$\tablenotemark[1]&-0.014&0.112&-0.021 \\
$\xi$            &-0.0099  &0.0558     &-0.0329 \\
$\tq_{\rm{A}}/e$ &26.843     &31.093     &4.961 \\
$\tq_{\rm{B}}/e$ &0.944      &0.385      &14.774 \\
$\tq_{\rm{I}}/e$ &-22.416  &-24.280  &-5.434 \\
$\tq_{\rm{II}}/e$ &-5.371  &-7.198   &-14.301 \\
$P^0\:$(C$\,$m$^{-2}$) &0.306  &-0.727  &0.358   \\
$|P^0_{\rm{exp}}|\,$(C$\:$m$^{-2}$)\tablenotemark[2]& 0.26
&0.75&0.37 \\
$|P^0_{\rm{BPA}}|\,$(C$\:$m$^{-2}$)&$0.30; 0.28$\tablenotemark[3]
$\hspace{-2.7em}$&$0.74; 1.04$\tablenotemark[4]$\hspace{-2.7em}$&
$0.35; 0.40$\tablenotemark[5]$\hspace{-2.7em}$
\end{tabular}
\end{ruledtabular}
\tablenotetext[1]{BaTiO$_3$ and PbTiO$_3$ in Ref. \cite{Wyck64},
KNbO$_3$ in Ref. \cite{Hewa73}.}
\tablenotetext[2]{BaTiO$_3$ in Ref. \cite{Merz53}, PbTiO$_3$ in
Ref. \cite{Gavr70}, KNbO$_3$ in Ref. \cite{Klee84}.}
\tablenotetext[3]{Ref. \cite{Zho94a} and Ref. \cite{Zho94b},
respectively.}
\tablenotetext[4]{Ref. \cite{Zho94a}, calculations at 700 K and at 295 K,
respectively!}
\tablenotetext[5]{Ref. \cite{Rest94} and Ref. \cite{Zho94a},
respectively.}
\end{table}
with treating $\Pb^0$ as an order parameter small near the transition
point \cite{Land84,Ginz46}. Furthermore, different nonzero projections of
$\Pb^0$ in (\ref{Eq20}) can arise separately, as it happens in BaTiO$_3$
and KNbO$_3$ \cite{Line77,Devo49,Slat50}.

Concentrating on regular states, here we estimate the polarization of
BaTiO$_3$, PbTiO$_3$, and KNbO$_3$ in a tetragonal phase. In general terms
of ABO$_3$ perovskite, such a ferroelectric phase is determined by the
lattice constants $a$ and $c$ and by extra displacements
$c\eta_{\mathrm{B}}$, $c\eta_{\mathrm{I}}$, and $c\eta_{\mathrm{II}}$ of
the B, O$_{\mathrm{I}}$, and two O$_{\mathrm{II}}$ sublattices relative to
the A one, provided that O$_{\mathrm{I}}$ and O$_{\mathrm{II}}$ ions
surround B along $c$ and in lateral directions, respectively, and positive
$\eta_{\mathrm{B}}$ defines the polar direction. The simple model of
uniform electron distributions within ionic spheres of Ba$^{2+}$,
Pb$^{2+}$, Ti$^{4+}$, K$^+$, Nb$^{5+}$, and O$^{2-}$ ions, with
O$_{\mathrm{II}}$ anions in contact, is employed. With making use of the
structural parameters listed in Table \ref{Table1} and with account of the
atomic numbers $Z_{\mathrm{Ba}}=56$, $Z_{\mathrm{Ti}}=22$,
$Z_{\mathrm{O}}=8$, $Z_{\mathrm{Pb}}=82$, $Z_{\mathrm{K}}=19$, and
$Z_{\mathrm{Nb}}=41$, the effective charges $\tq_j$ are obtained as a
solution of equations (\ref{Eq15})--(\ref{Eq19}) and are listed in Table
\ref{Table1} in units of the elementary charge $e$. Then the values of
$P^0$ are obtained with the help of (\ref{Eq13}), (\ref{Eq14}), and
(\ref{Eq20}) and agree with the experimental data, as shown in Table
\ref{Table1}. On comparing with the results of the Berry-phase approach
shown in Table \ref{Table1} as well, our estimate is much closer to the
experimental data on PbTiO$_3$ at room temperature \cite{Gavr70}, but is
comparable in other cases.

\end{document}